\documentclass[12pt]{article}
\usepackage{amsfonts}
\def\lp{\stackrel{\leftarrow}{\partial}}
\def\rp{\stackrel{\rightarrow}{\partial}}
\def\be{\begin{eqnarray}}
\def\ee{\end{eqnarray}}
\def\*{\star} 
\begin{document}
\centerline{\large AN ATAVISTIC LIE ALGEBRA} 

\phantom{aaa} 

\centerline{ \Large David B Fairlie$^{\sharp}$ and Cosmas K Zachos$^{\mho}$ }  

$^{\sharp}$Department of Mathematical Sciences, Durham University,
Durham, DH1 3LE, UK  \\
\phantom{a} \qquad   \qquad\qquad{\sl David.Fairlie@durham.ac.uk}

$^{\mho}$High Energy Physics Division,
Argonne National Laboratory, Argonne, IL 60439-4815, USA \\
\phantom{a}\qquad  \qquad \qquad  {\sl zachos@anl.gov}

\begin{abstract} 
An infinite-dimensional Lie Algebra is proposed which includes, in its
subalgebras and limits, most Lie Algebras routinely utilized in 
physics. It relies on the finite oscillator Lie group, and appears 
applicable to twisted noncommutative QFT and CFT.  
\end{abstract}
\bigskip

\hrule

\section{Introduction}

Consider the familiar forced oscillator 
\be
H = \alpha ^\dagger  ~\alpha  + \lambda (t)   \alpha^\dagger  
+\lambda^* (t)  \alpha   ~ ,   
\ee
driven by an external time-dependent source. 

Since $H=\exp ( -\lambda (t) \alpha^\dag + \lambda^*(t) \alpha ) ~ 
\Bigl( \alpha^\dag \alpha - |\lambda (t)|^2 \Bigr )~
\exp ( \lambda (t) \alpha^\dag - \lambda^*(t) \alpha) $, 
the integration of the model 
relies on the coherent states \cite{zhang} underlain by the noncompact 
oscillator group ${\cal G}$ (or ${\cal H}_4$) \cite{miller}.
This is the solvable, rank 2, dimension 4, Lie group generated by the 
oscillator creation and annihilation operators, their Heisenberg commutator
(which is central), and the occupation number operator. 
It turns out that this group, ${\cal G}$, as well as 
related Vertex algebras \cite{kac}, control  
the structure of an infinite dimensional Lie Algebra brought into play
in this letter.   
This algebra, remarkably, contains as subalgebras or limits of it, most 
of the Lie algebras routinely encountered in physics. 
 
The infinite dimensional Lie algebra we introduce in this letter and 
explore below is 
\be
[ J^a _{m_1,m_2}, J^b_{n_1,n_2}] =
e^{is(m_1 e^{-a}  n_2 - m_2 e^a n_1)} J^{a+b}_{m_1+e^a n_1, ~m_2+e^{-a} n_2}
-e^{is(n_1 e^{-b}m_2 - n_2 e^b m_1)} J^{a+b}_{n_1+e^b m_1, ~n_2+e^{-b} m_2} ~,
\label{Ogu}
\ee
where the upper indices, $a,b$, the lower ones, $m_1,m_2,..$ 
and the parameter $s$ are arbitrary, unless restricted by some further 
expediency.

It satisfies the Jacobi identity, which is evident from  its merely 
amounting to the antisymmetrization of the 
associative group product,
\be 
 J^a _{m_1,m_2} J^b_{n_1,n_2} =
e^{is(m_1 e^{-a}  n_2 - m_2 e^a n_1)} J^{a+b}_{m_1+e^a n_1, ~m_2+e^{-a} n_2}.
\label{product}
\ee
(Associativity amounts to $( J^a _{m_1,m_2} J^b_{n_1,n_2}) J^c _{k_1,k_2}=
J^a _{m_1,m_2} (J^b_{n_1,n_2} J^c _{k_1,k_2})$.)
Naturally, the symmetrization of this product into an anticommutator
further yields a consistent {\em graded extension} of the Lie algebra.

The algebra (\ref{Ogu}) contains, as a subalgebra specified by 
vanishing second subscripts,\newline 
$m_2=n_2=...=0$, the Vertex (coherent-state) Algebra \cite{vertex},
\be
[J^a_m , J^b_n ]  =  J^{a+b} _{m+ e^a n }  - J^{a+b}_{n + e^b m}~. \label{vrtx}
\ee 

Another, more familiar and general, subalgebra is the infinite Lie algebra 
specified by vanishing of all superscripts $a=b=...=0$, and by integer 
subscripts: it is the Sine Algebra \cite{ffz},
\be
[M_{m_1,m_2}, M_{n_1,n_2}] = 2i\sin\left ( s(m_1  n_2 - m_2  n_1) \right )~ 
M_{m_1+ n_1, m_2+n_2}~. \label{sinealg}
\ee
This one, in turn, for $s= -\hbar/2$, is recognized as the Moyal Bracket 
algebra on the basis of the Fourier modes $\exp(i m_1 x+im_2 p)$ of a 
toroidal phase space with unit radii \cite{ffz}. Up to a normalization, 
the Moyal Brackets represent the antisymmetrization of Groenewold's celebrated 
star product \cite{groenewold}, 
\be 
{\star \equiv e^{{i\hbar \over 2} (\lp_x \rp_p- \lp_p \rp_x )} } ~. 
\label{Ezili}  
\ee

Alternatively, for $s=2\pi/N$, and integer $N$, the sine algebra is seen
to represent $GL(N)$, and to thus include all classical Lie algebras, 
$(A_N,B_N, C_N, D_N)$ \cite{FFZjmp}. Furthermore \cite{ffz},  for 
$s\rightarrow 0$  (or $N\rightarrow \infty$), this algebra goes to  
the algebra of Poisson Brackets, also realized on a 
toroidal phase space, SDiff$(T^2)$, cf \cite{antoniadis};  
and this one contains the Virasoro algebra as a subalgebra, 
through judicious summation over the first
subscripts $m_1, n_1,...$,  \cite{antoniadis}. 

This, then, is the basis of our remark that the Lie Algebra introduced 
encompasses and frames quite a range of the Lie algebras normally 
utilized in physics.

In this letter, we explore basic features of this Lie algebra, eqn 
(\ref{Ogu}),
its applicability, and generalization. We provide a few explicit useful
realizations and representations of restrictions thereof, which illuminate 
its structure. 

Throughout, we stress the
somewhat untypical correspondence mentioned, which applies both to the algebra
(\ref{Ogu}) itself, as well as the subalgebras we sketched: namely, a
{\em finite-dimensional Lie group} product,  which yields the {\em infinite
Lie algebras} when antisymmetrized---and graded (anticommutator) extensions,
when symmetrized.

\section{The Atavistic Algebra, realizations, and representations} 
The Atavistic Algebra (\ref{Ogu}) product (\ref{product}) might be rewritten
more symmetrically, through the redefinition
\be
V^a_{m_1,m_2} \equiv  J^{2a}_{e^a m_1  , ~  e^{-a} m_2}, \label{Vsymm}
\ee
so that the product (\ref{product}) amounts to
\be
  V^a_{m_1,m_2}~ V^b_{n_1,n_2}= e^{is(m_1 n_2 e^{-a-b}-m_2 n_1 e^{a+b})} ~
V^{a+b}_{e^{-b} m_1+e^a n_1, ~e^b m_2+ e^{-a} n_2} .
\ee
Nevertheless, we stick to the original form, to minimize superfluous indices.
 
The semidirect product nature of eqn (\ref{product}) and thus of the 
Atavistic Algebra (\ref{Ogu}) is readily recognizable by recalling the 
$\star$-product 
underlying the sine algebra (\ref{sinealg}). We first revert to 
the version of the star product  (\ref{Ezili}) parameterized for our 
purpose,
\be 
{\star \equiv e^{-i s (\lp_x \rp_p- \lp_p \rp_x )} } ~, 
\ee
where $x,p$ range from 0 to 1.  
The most widely used property of this product is its associativity. 
Thus, strings of operators of the form   $f(x,p)\star $, for any functions
$f(x,p)$ on the unit $T^2$, may be equivalently evaluated indifferently to 
the grouping of multiplication chosen. Consequently, choosing a Fourier 
mode basis for integers $m_1, m_2$, and defining 
\be
M_{m_1,m_2}\equiv \exp (i(m_1 x + m_2 p)) ~ \star  ~,
\ee
 the standard product law follows \cite{ffz},   
\be
M_{m_1,m_2} M_{n_1,n_2} = \exp \left ( is(m_1  n_2 - m_2  n_1) \right )~ 
M_{m_1+ n_1, m_2+n_2}~,  \label{Agwe}
\ee
underlying eqn (\ref{sinealg}) when antisymmetrized. The finite Lie group 
corresponding to this product is well-known to be the (dimension 3) 
Heisenberg group \cite{weyl}.

Now consider a phase-space-area-preserving dilation operator $D(a)$,
which braids associatively as  
\be
D(a) ~f(x,p) = f(e^a x, e^{-a} p) ~D(a).
\ee 
A standard realization of this operator is 
\be
D(a) = \exp \left ( a (x\rp_x -p\rp_p ) \right ).
\ee

Moreover, this operator formally commutes with the above star product,
\be
D(a) ~\star  =  \star  ~D(a) ,
\ee
as is plain in integral kernel representations
of the star product \cite{groenewold}. 
More directly, this also follows from braiding $D(a)$  past the mode
\be 
\left ( e^{i(m_1 x+m_2 p)} \star e^{i(m_1 x+m_2 p)}\right ) =
e^{is(m_1 n_2-m_2 n_1)}~ e^{i((m_1+n_1)x+(m_2+n_2)p)},
\ee 
as above, eqn (\ref {Agwe}). Braiding past the simple right-hand side
 must equal stepwise braiding past each of the factors on the $\star$-product 
on the left, so it is manifest that the $\star$-product remains invariant.
Of course,  the reason for the invariance is the area-preservation feature of 
the dilation operator chosen, since the phase proportioned by $s$ is a 
two-dimensional cross product, amounting to a phase-space-area element.

Furthermore, evidently, 
\be
D(a)D(b)=D(a+b),    \qquad \qquad \qquad  D(0)= 1\kern -0.36em\llap~1 .
\ee

Consequently, it follows directly that the Atavistic Algebra elements  
$J^a _{m_1,m_2}$
may be constructed out of $ M_{m_1,m_2} D(a)=  J^0_{m_1,m_2}D(a)$, i.e.,
\be
J^a _{m_1,m_2} = \exp (i(m_1 x + m_2 p)) ~ \star  ~D(a).
\ee
Writing this out explicitly,
\be
J^a _{m_1,m_2} = e^{i(m_1 x + m_2 p)}  ~  e^{s(m_1 \partial _p  - 
m_2 \partial_x)} ~  e^{ a (x\partial_x -p\partial_p) }  ,
\ee
the reader may recognize that the product (\ref{product}) and thus 
the Atavistic Algebra (\ref{Ogu}) are, indeed, satisfied.    

Thus, operation by the $J^a _{m_1,m_2}$ on a function $f(x,p)$ on this 
phase space consists of sequential rescalings and shifts of its variables
and multiplication by a phase,
\be
J^a _{m_1,m_2} f(x,p) = e^{i(m_1 x + m_2 p)}~f\Bigl (e^{a}(x-sm_2), 
e^{-a}(p+sm_1)\Bigr ) .
\label{functionaction}
\ee

The form (\ref{Vsymm}) corresponds to $V^a_{m_1,m_2}= D(a) J^a _{m_1,m_2}$.

\section{Reduction of variables}
Now note that the torus variables $x,p$ commute with each other, so 
that, effectively, the above realization is a direct product 
of the same type of operator
\be
{\mathfrak J} ^a_{m_1,m_2} \equiv  e^{i s m_1 m_2/2 } ~
 e^{i m_1 x}  ~  e^{ s  m_2 \partial_x}
 ~  e^{ a x\partial_x} , 
\ee
acting on variables $x$, and, disjointly, acting on variables $p$,
which is to say
\be
J^a_{m_1,m_2} = {\mathfrak J}^a_{m_1,-m_2} \otimes  {\mathfrak J}^{-a}_{m_2,m_1} .
\ee

In turn, the product of these operators is 
\be 
 {\mathfrak J}^a _{m_1,m_2} {\mathfrak J}^b_{n_1,n_2} =
e^{-is(m_1 e^{-a}  n_2 - m_2 e^a n_1)/2} ~ 
{\mathfrak J}^{a+b}_{m_1+e^a n_1,~m_2+e^{-a} n_2}~,
\ee
i.e., {\em the same as eqn (\ref{product}), for minus half the value of the 
parameter} $s$.
It is plain that, in general (irrespective of realization),
\be
J^a_{m_1,m_2} (s) =~ J^a_{m_1,-m_2}(s/2)  \otimes  ~J^{-a}_{m_2,m_1}(s/2) ~ .
\ee
This halving of representations and phases is already apparent in 
the first of refs \cite{ffz}: a realization introduced by Hoppe there 
effectively represents the star product on {\em half the variables} of 
phase space. 
Plainly, this recursive phase-addition phenomenon extends to arbitrarily 
long direct product strings of $J$s, as a reflection of the simple 
coproduct structure of their logarithms\footnote{The mathematically 
inclined reader may recognize the connection of this relation
to the Drinfeld twist of quantum deformation coproducts, as implemented by 
Reshetikhin \cite{reshetikhin}, and exemplified, e.g., by ref \cite{cgz}, 
eqn (2.13), or ref \cite{chaichian}, eqn (2.5), etc. These outrange the scope 
of the present discussion.}.

Equivalently,  since oscillator operators, $[\alpha ,\alpha ^\dag ]=1$,
formally  parallel the commutation relations of $[\partial_x, x] =1$, 
the above realization may also be displayed in a ``coherent state" form,
\be
{\mathfrak J}^a_{m_1,m_2} =  
 e^{i m_1 \alpha^\dag +s  m_2 \alpha }
 ~  e^{ a~ \alpha ^\dag \alpha } .  \label{Attunement}
\ee
In this form, the general Lie subalgebras specified by all 
$m_2\rightarrow 0$, yielding  (\ref{sinealg}); and $a\rightarrow 0$,
yielding (\ref{vrtx}), are manifest.

One would not expect finite dimensional representations for the generic 
Atavistic Algebra. 
If the expression eqn (\ref{Attunement}) looks familiar, it is because it is.
It is, in fact, the parameterization of the oscillator Lie 
group ${\cal G}$ investigated by Miller \cite{miller}. He provides, in 
gratifying detail, the representations
of this Lie group on finite vector spaces (by matrices), unfaithfully; 
and, faithfully, on infinite-dimensional Bargmann space, by differential 
operators, eqn (\ref{functionaction})---for instance, since 
${\mathfrak J}^a_{m_1,m_2} f(x)=e^{i m_1 (x-sm_2/2)}$ $f(e^{a} (x+ s  m_2) )$, 
representations on associated Laguerre polynomials  are specified by this 
simple action on the generating functions of these polynomials. 
Representations on Hermite Hilbert space then follow \cite{miller}.  

The four generators of the oscillator group ${\cal G}$
are abstractions of the operators 
$\alpha ,\alpha^\dag $, $1\kern -0.36em\llap~1$,
augmented by the occupation number operator, normally 
$\alpha^\dag \alpha$. These are now abstracted to 
\be
[{\cal A} , {\cal A}^\dag ]={\cal E}, \quad
[{\cal N}, {\cal A}]= -{\cal A}, \quad
[{\cal N}, {\cal A}^\dag]= {\cal A} ^\dag, \quad  
[{\cal E}, {\cal N}] = [{\cal E}, {\cal A}]=  [{\cal E}, {\cal A}^\dag]= 0.
\label{FlimaniKoku}
\ee
Faithful representations of this  need be infinite dimensional;
but unfaithful ones could be finite dimensional, as long as 
the central generator ${\cal E}$ is not the identity.
This can be arranged by nilpotent matrices.

For example \cite{miller}, one might consider 3$\times$3 nonhermitean matrices,
\begin{equation}
{\cal A} = \left( \begin{array}{ccc}
          0&0&0  \\
          0&0&1  \\
          0&0&0 
            \end{array} \right)~,
\quad {\cal A}^\dag = 
\left( \begin{array}{ccc}
          0&1&0 \\
          0&0&0 \\
          0&0&0 
\end{array} \right)~, 
\quad {\cal N} = 
\left( \begin{array}{ccc}
          1&0&0 \\
          0&0&0 \\
          0&0&1 
\end{array} \right)~, 
\quad {\cal E} = 
\left( \begin{array}{ccc}
          0&0&-1 \\
          0&0&0 \\
          0&0&0 
\end{array} \right)~.
\end{equation}
These represent eqn (\ref{FlimaniKoku}), although note that 
${\cal A}^\dag {\cal A} =-{\cal E}\neq {\cal N}$, ${\cal N}^2={\cal N}$, 
and $({\cal A}^\dag) ^2= {\cal A}^2 ={\cal E}^2=0$.
In this representation, the proper abstracted analog of (\ref{Attunement}) reads
\begin{equation}
{\cal J}^a_{m_1,m_2} = 
\left( \begin{array}{ccc}
          e^a  &im_1  & -is m_1m_2 e^a /2 \\
          0&1&sm_2e^a   \\
          0&0&e^a  
            \end{array} \right)~,
\end{equation}
and satisfies, instead, 
\be 
 {\cal J}^a _{m_1,m_2} {\cal J}^b_{n_1,n_2} =
\exp\!\left ( {-is\over 2} (m_1 e^{-a}  n_2 - m_2 e^a n_1) {\cal E}\right ) ~
{\cal J}^{a+b}_{m_1+e^a n_1, ~m_2+e^{-a} n_2}~.
\ee
However,
\be
\exp \left ( -{is\over 2}(m_1 e^{-a}  n_2 - m_2 e^a n_1) {\cal E}\right ) =
1\kern -0.36em\llap~1 -{is\over 2}(m_1 e^{-a}  n_2 - m_2 e^a n_1) {\cal E}
\ee
is not a pure phase, as the (traceless) center ${\cal E}$ is not the 
identity matrix. 

The inadequacy of this construction in representing the pure phase appears generic
to all finite dimensional representations sought. Nevertheless, we illustrate
explicitly  in the next section 
that finite dimensional matrix representations can be available for
suitable {\em restrictions of the Atavistic Algebra}.

\section{Finite matrix representations for restricted cases}

Suppose, instead, that we restrict the parameters and 
indices of the product (\ref{product}) to $s=-\pi/p$ for an odd prime 
integer $p$,
so that $\exp(-2is)\equiv \omega=e^{2\pi i/p}$, with $\omega^p=1$;
and take integer subscripts mod $p$, $m_j=0,1,2,...,p-1$;
and rescaled superscripts to be integer mod $p-1$,  
$\tilde{a}  \equiv a/\ln 2  = 0,1,2,...,p-2$, recalling cyclicity:
$2^{p-1}=1$ mod $p$, for any odd prime integer $p$.

The product now reads
\be 
J^{\tilde{a}}   _{m_1,m_2} J^{\tilde{b}}  _{n_1,n_2} =
\omega^{(2^{\tilde{a}}   m_2 n_1 - 2^{p-1-\tilde{a}  } m_1 n_2)/2 }  ~
J^{\tilde{a}+\tilde{b}}_{m_1+2^{\tilde{a}}n_1, ~m_2+2^{p-1-\tilde{a}} n_2}. 
\label{Legba}
\ee 

This product (and the corresponding antisymmetrization Lie Algebra) can 
be represented by Sylvester's celebrated $p\times p$ 
matrix basis for $GL(p)$ groups \cite{sylvester}. His 
standard clock and shift unitary unimodular matrices,
are 
\be
Q_{rt}=\omega^r \delta_{r,t}~, \qquad \qquad 
P_{rt}= \delta_{r+1,t}~, \qquad \qquad 
\ee
for indices $r,t$ defined mod $p$, $r=0,1,2,...,p-1$.

Consequently, 
\be
Q^p=P^p= 1\kern -0.36em\llap~1 , \qquad \qquad  PQ =\omega ~ QP,    
\ee
the characteristic braiding identity  \cite{weyl,ffz,FFZjmp}.

The complete set of 
$p^2$ unitary unimodular $p\times p$ matrices
\be
M_{(m_1,m_2)}
\equiv \omega ^{m_1 m_2 /2}\, Q^{m_1} P^{m_2},
\qquad \Longrightarrow \qquad 
   (M_{(m_1,m_2)})_{rt}   =\omega^{m_1 (r+m_2 /2)} \delta_{r+m_2,t}~, 
\ee
where 
\be
M^{\dag}_{(m_1,m_2)}    =M_{(-m_1,-m_2)},
\ee
and Tr$M_{(m_1,m_2)}= p \delta_{m_1,0} \delta_{m_2,0}    $,
suffice to span the group algebra of $GL(p)$. Since
\be
M_{(m_1,m_2)}    M_{(n_1,n_2)}    =
\omega^{(m_2 n_1 - m_1 n_2)/2 }  ~ M_{(m_1+n_1,m_2+ n_2)}~,
\ee
they further satisfy the Lie algebra of $SU(p)$, \cite{ffz},  
a restriction of the Sine Algebra displayed in the Introduction,
\be
[M_{(m_1,m_2)} ,   M_{(n_1,n_2)} ]   =
2i~\sin\left ( {\pi\over p} (m_2 n_1 - m_1 n_2)  \right )
 M_{(m_1+n_1,m_2+n_2)} ~.
\ee

Now, in addition, consider the discrete scaling (doubling) matrix 
\cite{vourdas},
\be
R_{rt}\equiv \delta_{2r,t}~,  \qquad \qquad  R^{p-1}= 1\kern -0.36em\llap~1 ,
\qquad \qquad  R^\dagger=R^T= R^{p-2},
\ee
again for indices $r,t$ defined mod $p$, $r=0,1,2,...,p-1$.
The cyclic structure holds by virtue of the identity $2^{p-1}=1$ mod $p$.
(Note that a smaller power of 2 may return to 1 before the power $p-1$, e.g. 
$2^3=1$ mod 7, but this does not compromise the present construction.)

The action of the doubling matrix is
\be
R Q R^{p-2} = Q^2, \qquad \qquad  R^{p-2}  P R = P^2.
\ee
Consequently,
\be 
R^{\tilde{a}}   Q^{m_1} P^{m_2} R^{p-1-\tilde{a}}= 
Q^{ 2^{\tilde{a}} m_1} P^{2^{p-1-{\tilde{a}}} m_2}, 
\ee
and hence the unitary matrices 
\be
{\cal J}^{\tilde{a}}  _{m_1,m_2} \equiv  M_{(m_1,m_2)}
   R^{\tilde{a}}   
\ee
satisfy eqn (\ref{Legba}). 
Thus, these ${\cal J}$s provide a $p$-dimensional representation, and hence 
a representation of the restricted Atavistic Lie Algebra.
Through the direct product recursive procedure of the preceding section,
they can be augmented to selected higher dimensional ones.

However, since Sylvester's basis is complete, $R$ is representable in terms of 
the above $p^2~$  $M$s---in fact, it is the phased sum of all $~p\times p$ 
matrices $M$, normalized by $p$, since, $\forall m_1,m_2$, 
\be
\hbox{Tr} ~ M_{(m_1,m_2)} R=  \omega^{-3 m_1 m_2/2}~. 
\ee
Thus, since, e.g., 
\be
p {\cal J}^1_{0,0}-\sum_{m_1,m_2} \omega^{-3 m_1 m_2/2}~{\cal J}^0_{m_1,m_2}
=0 ~,
\ee
is represented trivially, the representations displayed are not faithful.

It is not hard to generalize to scaling (squeezing) matrices 
\cite{vourdas} with analogous effects.
For instance, ``Fermat's Little Theorem" \cite{Weisstein} dictates
$n^{p-1}=1$ mod $p$, for any odd prime $p$ and, e.g., $0<n<p$. Thus,  
instead of $R$, a matrix $T$  scaling  by  $n$,
\be
T_{rt}\equiv \delta_{n r,t}~,  \qquad \qquad  T^{p-1}
= 1\kern -0.36em\llap~1 ,
\qquad \qquad  T^\dagger=T^T= T^{p-2},
\ee 
acts  through log base  $n$, i.e. $a= \tilde{a}   \ln n$,
\be 
T^{\tilde{a}}   Q^{m_1} P^{m_2} T^{p-1 -\tilde{a}  }= Q^{ n^{\tilde{a}} 
m_1} P^{n^{p-1 -\tilde{a}  } m_2}, 
\ee
yielding analogous representations for the restricted Atavistic Lie Algebra.  
The question as to whether suitable tensor products of such spaces could
yield faithful representations, instead, remains open.

\section{Generalizations to linear canonical transforms\\
 and discussion}

It turns out that $D(a)$ in  Section  2 need not be so restricted,
if an associative product of the above type is sought. 
Indeed, any $Sp(2)$ linear symplectic transformation will do,
when it comes to leaving the star product invariant \cite{kim} 
(over and above preserving areas and Poisson Brackets---but nonlinear 
SDiff$(T^2)$ does not, in general). 
Thus, $D(a)$ may be generalized to a matrix ${\bf S}$ with $\det {\bf S}=1$, in
\begin{equation}
\left( \begin{array}{c}
          x \\
          p 
            \end{array} \right )~~ \mapsto ~~{\bf S}~ 
\left( \begin{array}{c}
          x \\
          p
            \end{array} \right)~.
\end{equation}
The above $D(a)$ corresponds to the special case ${\bf S}=$diag$(e^a,e^{-a})$. 
The transformation matrix needs transposition, ${\bf S}^T$, to act on the 
Fourier mode coefficients' doublets $(m_1,m_2)$. The Atavistic Algebra then 
further extends to the associative products of operators 
\be
J^{\bf S} _{m_1,m_2} = \exp (i(m_1 x + m_2 p)) ~ \star  ~{\bf S},
\ee
where, naturally, the matrix superscripts multiply, and the cross product in
the phase is between the untransformed and suitably transformed subscript 
doublet. This formal structure parallels symplectic squeezing of light 
\cite{zhang,kim,vourdas}, where the $Sp(2)$ linear symplectic transformation 
amounts to a Bogoliubov transformation.

We expect that this product structure, whose antisymmetrization yields the
corresponding generalized Atavistic Lie Algebra, is sufficiently general to
find applications in a broad variety of physics contexts. That is, the
associative product $f(x,p) \star g({\bf S}(x,p) )$ is a tractable
generalization of the standard $\star$-product, and should be relevant
to squeezed state, or lattice/brane deconstruction contexts, in which
$f(x,p)$ is defined at one end of a link (brane) and $g(x,p)$ on the other,
the symplectic transformation  ${\bf S}$ providing the link transition
function. Moreover, the Drinfeld twist coproduct exemplified in Section 3
bears compelling intuitive connections to applications of this twist in
noncommutative QFT \cite{chaichian} and deformation-generalized CFT
\cite{matlock}, currently under investigation.

\noindent{\it This work was supported by the US Department of Energy, 
Division of High Energy Physics, Contract W-31-109-ENG-38; and the 
Collaborative Project GEP1-3327-TB-03 of the US Civilian Research
and Development Foundation. 
Helpful discussions with A Vourdas, Y Meurice and G Jorjadze, are 
gratefully acknowledged.}

\end{document}